\def\today{\ifcase\month\or
  January\or February\or March\or April\or May\or June\or
  July\or August\or September\or October\or November\or December\fi

  \space\number\day, \number\year}

\hfuzz 30pt

\catcode`\:=11
%%%%%%%%%%%%%%%%%%%%%%%%%%%%%%%%%%%%%%%%%%%%%%%%%%%%%%%%%%%%%%%%%%%%%%
%COUNTERS, REGISTERS AND PRELIMINARY DEFINITIONS
%%%%%%%%%%%%%%%%%%%%%%%%%%%%%%%%%%%%%%%%%%%%%%%%%%%%%%%%%%%%%%%%%%%%%%
\newcount\bm:counta \newcount\bm:countb 
\newcount\bm:countc \newcount\bm:countd
\newtoks\bm:tok
\newif\ifbm:delim
\def\thehex#1{\ifcase\the#1 0\or 1\or 2\or 3\or 4\or 5\or 6\or 7\or
8\or 9\or A \or B\or C\or D\or E\or F\fi}%
\def\test#1#2{\ifcat#1#2\message{True}\else\message{False}\fi}
\newtoks \bm:savedtoks  \bm:savedtoks{}%
\def\bm:empty{\relax}%
\let\bm:save=\bm:empty
\newif\ifbm:cdr
\def\bm:split#1#2\bm:empty{%
    \def\bm:car{#1}\def\bm:cdr{#2\relax}%
    \expandafter\ifx\expandafter\relax\bm:cdr\bm:cdrfalse
    \else\bm:cdrtrue
    \fi}%
\newif\ifbm:found
\def\bm:in#1\find:#2\this:{%
    \def\find:##1#2##2##3\find:{%
        \ifx\bm:in##2\bm:foundfalse
        \else\bm:foundtrue
        \fi}%
    \find:#1#2\bm:in\find:}%
\let\when=\relax \let\use=\relax
%
%
%WARNINGS
\newif\ifboldwarning
\def\bm:message#1{{\newlinechar=`^^J
\immediate\write16{\string\bold\space warning on line
                                             \the\inputlineno^^J#1^^J}}}%
%
%LATEX FONT SELECTION ADJUSTMENT
\def\latex:adjust{\expandafter\ifx\the\textfont0\csname twlrm\endcsname
                                           \def\bm:scale{1200}%
                              \else\expandafter\ifx
                                 \the\textfont0\csname elvrm\endcsname
                                                 \def\bm:scale{1095}%
                                               \else\def\bm:scale{1000}%
                                               \fi
                              \fi}%
\latex:adjust
\newdimen\bm:sevensize \newdimen\bm:fivesize
\bm:sevensize=.007pt \bm:fivesize=.005pt
\bm:sevensize=\bm:scale\bm:sevensize
\bm:fivesize=\bm:scale\bm:fivesize
%BOLD ROMAN FAMILY (REDEFINITIONS NECESSARY FOR LATEX!)
\font\tenbf=cmbx10 scaled \bm:scale
\font\sevenbf=cmbx7 at \the\bm:sevensize
\font\fivebf=cmbx5 at \the\bm:fivesize
\textfont\bffam=\tenbf
\scriptfont\bffam=\sevenbf
\scriptscriptfont\bffam=\fivebf
%BOLD TEXT ITALIC FAMILY
:bit=cmbxti10 scaled \bm:scale
:bit=cmbxti10 at \the\bm:sevensize
:bit=cmbxti10 at \the\bm:fivesize
%\newfam\bitfam
%\textfont\bitfam=\tenbm:bit
%\scriptfont\bitfam=\sevenbm:bit
%\scriptscriptfont\bitfam=\fivebm:bit
%SANS FAMILY
:scale
 at \the\bm:sevensize
 at \the\bm:fivesize
%\newfam\sffam
%\textfont\sffam=\tensf scaled \bm:scale
%\scriptfont\sffam=\sevensf
%\scriptscriptfont\sffam=\fivesf
%BOLD SANS FAMILY
:bsf=cmssbx10 scaled \bm:scale
:bsf=cmssbx10 at \the\bm:sevensize
:bsf=cmssbx10 at \the\bm:fivesize
%\newfam\bsffam
%\textfont\bsffam=\tenbm:bsf
%\scriptfont\bsffam=\sevenbm:bsf
%\scriptscriptfont\bsffam=\fivebm:bsf
%BOLD SLANTED FAMILY
:bsl=cmbxsl10 scaled \bm:scale
:bsl=cmbxsl10 at \the\bm:sevensize
:bsl=cmbxsl10 at \the\bm:fivesize
%\newfam\bslfam
%\textfont\bslfam=\tenbm:bsl
%\scriptfont\bslfam=\sevenbm:bsl
%\scriptscriptfont\bslfam=\fivebm:bsl
%BOLD MATH ITALIC FAMILY
:bmit=cmmib10 scaled \bm:scale
:bmit=cmmib7 at \the\bm:sevensize
:bmit=cmmib5 at \the\bm:fivesize
\newfam\bm:bmitfam
\textfont\bm:bmitfam=\tenbm:bmit
\scriptfont\bm:bmitfam=\sevenbm:bmit
\scriptscriptfont\bm:bmitfam=\fivebm:bmit
%BOLD SYMBOL FAMILY
:bsy=cmbsy10 scaled \bm:scale
:bsy=cmbsy7 at \the\bm:sevensize
:bsy=cmbsy5 at \the\bm:fivesize
\newfam\bm:bsyfam
\textfont\bm:bsyfam=\tenbm:bsy
\scriptfont\bm:bsyfam=\sevenbm:bsy
\scriptscriptfont\bm:bsyfam=\fivebm:bsy
%FAMILY ASSIGNMENT FOR PURELY `ALPHABETIC' FONTS
\newtoks\alphatok \alphatok{?}%
\expandafter\def\expandafter\new:fam\expandafter{\newfam}%
\def\set:fam#1{%
    \expandafter\ifx\csname #1fam\endcsname\relax
                    \expandafter\new:fam\csname #1fam\endcsname
                    \edef\alphafam{\csname #1fam\endcsname}%
                    \edef\set:fonts{% 
                         \global\textfont\alphafam=\csname ten#1\endcsname
                         \global\scriptfont\alphafam=\csname
                                                seven#1\endcsname 
                         \global\scriptscriptfont\alphafam=\csname
                                                        five#1\endcsname}%
                    \set:fonts
                \fi
}%
\def\declare:alpha#1{%
    \set:fam{#1}%
    \expandafter\edef\csname math#1\endcsname{{%
                     \noexpand\if?\noexpand\the\noexpand\alphatok
                                  \global\noexpand\bm:savedtoks
                            {\noexpand\csname math:#1\noexpand\endcsname}%
                                  \noexpand\aftergroup\noexpand\bm:getarg
                     \noexpand\else\errmessage{You're already inside a
                                   \noexpand\expandafter\noexpand\string
                                   \noexpand\the\alphatok{..} - don't
                                   even think about it!}%
                     \noexpand\fi}}%
    \expandafter\def\csname math:#1\endcsname##1{{%
                     \alphatok\expandafter{\csname math#1\endcsname}%
                     \fam\csname #1fam\endcsname
% DISABLED TO ALLOW \ \csname ten#1\endcsname
                     \let\boldletter=\relax ##1}}}%
%
%FONT COMMANDS
\chardef\rmfam=0
\chardef\mitfam=1
\chardef\calfam=2
\declare:alpha{rm}%
\declare:alpha{it}%
\declare:alpha{sl}%
\declare:alpha{tt}%
\declare:alpha{bf}%
\declare:alpha{mit}%
\declare:alpha{cal}%
%%%%%%%%%%%%%%%%%%%%%%%%%%%%%%%%%%%%%%%%%%%%%%%%%%%%%%%%%%%%%%%%%%%%%%%%%%
\declare:alpha{sf}%%%%%%%%%%%%%%%%%%%%%%%%%%%%%%%%%%%%%%%%%%%%%%%%%%%%%%%%
\declare:alpha{bm:bmit}%%%%%%%%%%%%%%%%%%%%%%%%%%%%%%%%%%%%%%%%%%%%%%%%%%%
\declare:alpha{bm:bsy}%%%%%%%%%%%%%%%%%%%%%%%%%%%%%%%%%%%%%%%%%%%%%%%%%%%%
\declare:alpha{bm:bsf}%%%%%%%%%%%%%%%%%%%%%%%%%%%%%%%%%%%%%NEW PRIVATE  %%
\declare:alpha{bm:bit}%%%%%%%%%%%%%%%%%%%%%%%%%%%%%%%%%%%%%ALPHABETIC   %%
%\declare:alpha{bm:bsl}%%%%%%%%%%%%%%%%%%%%%%%%%%%%%%%%%%%%%FONT COMMANDS%%
%%%%%%%%%%%%%%%%%%%%%%%%%%%%%%%%%%%%%%%%%%%%%%%%%%%%%%%%%%%%%%%%%%%%%%%%%%
\def\default{\fam=-1 \def\boldletter##1{{\bf ##1}}}%
%%%%%%%%%%%%%%%%%%%%%%%%%%%%%%%%%%%%%%%%%%%%%%%%%%%%%%%%%%%%%%%%%%%%%%%
%SHORT ALIASES FOR FONT COMMANDS
\let\mathbm=\mathbm:bmit :bsy
\let\mathbm:bcal=\mathbm:bsy
\def\bold#1{{% begin group
    \let\bm:currentsymbol=\relax
    \bm:split#1\bm:empty
    \ifbm:cdr\toks0{}\loop\toks0\expandafter\expandafter\expandafter{\expandafter\the\expandafter\toks0\expandafter\noexpand\expandafter\bold\expandafter{\bm:car}}\expandafter\bm:split\bm:cdr\bm:empty
                      \ifbm:cdr
                      \repeat
              \toks2\expandafter{\bm:car}%
              \xdef\bm:out{\the\toks0
                           \noexpand\bold\expandafter{\the\toks2}}%
              \aftergroup\bm:out 
    \else\ifx\bold#1\aftergroup\bold
         \else\ifmmode\bm:select{#1}%
                      \ifx\bm:save\bm:empty
                          \xdef\bm:out{\global\bm:savedtoks{}%
                             \the\bm:savedtoks\noexpand\bm:currentsymbol}%
                          \aftergroup\bm:out
                      \else\global\bm:savedtoks\expandafter\expandafter
          \expandafter{\expandafter\the\expandafter\bm:savedtoks\bm:save}%
                            \ifx\bm:currentsymbol\relax\aftergroup\bold
                            \else\let\test=F%
                                 \edef\argtest{\noexpand\bm:in
                                      \meaning\bm:currentsymbol
                                      \noexpand\find:
                                      \string\mathaccent
                                      \noexpand\this:}%
                                 \argtest
                                 \ifbm:found \let\test=T%
                                 \fi
                                 \edef\argtest{\noexpand\bm:in
                                      \meaning\bm:currentsymbol
                                      \noexpand\find:
                                      \string\radical
                                      \noexpand\this:}%
                                 \argtest
                                 \ifbm:found \let\test=T%
                                 \fi
                                 \if T\test\aftergroup\bm:getarg
                                 \else\aftergroup\bm:dumpchars
                                 \fi
                            \fi
                      \fi
              \else\errmessage{\string\bold\space should be used in
                               math mode only}%
              \fi
          \fi
    \fi
    }}%
\def\bm:getarg#1{{%
    \ifx#1\bold\aftergroup\bold
    \else\toks0{#1}\xdef\bm:out{\global\bm:savedtoks{}%
                        \the\bm:savedtoks{\the\toks0}}%
         \aftergroup\bm:out
    \fi}}%
\def\bm:dumpchars{{%
    \xdef\bm:out{\global\bm:savedtoks{}\the\bm:savedtoks}%
    \aftergroup\bm:out}}%
\def\bm:select#1{%
    \expandafter\bm:in\boldspecials\find:#1\this:
    \ifbm:found \def\when##1\use##2;{%
                    \ifx##1#1\xdef\bm:currentsymbol{\noexpand##2}%
                    \else\ifx##2#1\xdef\bm:currentsymbol{\noexpand##2}\fi
                    \fi}%
                \boldspecials
    \else\if\noexpand#1\relax                    %IT'S A CONTROL SEQUENCE
            \let\test=F%
            \edef\chartest{\noexpand\bm:in\meaning#1\noexpand\find:
                                         \string\mathchar\noexpand\this:}%
            \chartest
            \ifbm:found\let\test=T%
            \fi
            \edef\chartest{\noexpand\bm:in\meaning#1\noexpand\find:
                                           \string\char\noexpand\this:}%
            \chartest
            \ifbm:found \let\test=T%
            \fi                                  %IT'S A CHAR OR MATHCHAR
            \if T\test 
          \expandafter\ifx\csname bold\string#1\endcsname\relax 
                          \edef\bm:process{\noexpand\defboldsymbol
                                           {\noexpand#1}%
                                           {\noexpand\mathchar}%
                                           {\the#1}{-1}}%
                          \bm:process
                       \fi \global\expandafter\let\expandafter
                           \bm:currentsymbol\expandafter=%
                           \csname bold\string#1\endcsname
            \else %IF NOT A CHAR OR MATHCHAR SAVE IT TO DELAY EXPANSION
          \expandafter\ifx\csname bold\string#1\endcsname\relax 
                          \ifboldwarning\bm:message{Skipping \string#1.}%
                          \fi
                                        \def\bm:save{#1}%
                      \else\toks4\expandafter{%
                           \csname bold\string#1\endcsname}%
                           \edef\bm:save{\the\toks4}%
                           \global\expandafter\let\expandafter
                           \bm:currentsymbol\expandafter=%
                           \the\toks4
                      \fi
            \fi
         \else                                   %IT'S A CHAR TOKEN OR
						 %SOMETHING TO SKIP
              \ifcat\noexpand#1A\gdef\bm:currentsymbol{\boldletter{#1}}%
              \else\ifcat\noexpand#1>            %IT'S A NONLETTER
             \expandafter\ifx\csname bold\string#1\endcsname\relax
                             \edef\bm:process{\noexpand\defboldsymbol
                                              {\noexpand#1}%
                                              {\ifnum\the\delcode`#1>-1
                                                 \noexpand\delimiter
                                               \else\noexpand\mathchar
                                               \fi}%
                                              {\the\mathcode`#1}%
                                              {\the\delcode`#1}}%
                             \bm:process         %DEFINE BOLD SYMBOL
						 %IF NECESSARY
                          \fi\global\expandafter\let
                             \expandafter\bm:currentsymbol
                             \expandafter=\csname 
                             bold\string#1\endcsname
                   \else\ifboldwarning\bm:message{Skipping \string#1.}%
                        \fi
                                      \def\bm:save{#1}%
                   \fi
              \fi
         \fi
    \fi}%
\def\defboldsymbol#1#2#3#4{%
    \bm:tok={#1}%
 \expandafter\ifx\csname bold\string#1\endcsname\relax
             \else\ifboldwarning\bm:message{Redefining
                                                \string\bold\string#1.}%
                  \fi
             \fi
    \bm:counta=#3
    \bm:countd=#4
    \ifnum\the\bm:counta="8000
           \expandafter\xdef\csname bold\string#1\endcsname{#1}%
    \else\ifx#2\delimiter \bm:delimtrue
         \else\ifx#2\radical \bm:delimtrue
              \else \bm:delimfalse
              \fi
         \fi
         \bm:countc=\bm:counta
         \divide\bm:countc by "1000             %C HOLDS THE CLASS DIGIT
         \advance\bm:counta by -\expandafter"\thehex\bm:countc 000
                                                % SINGLE DIGIT<16
         \ifbm:delim\ifnum\the\bm:countd>-1     
                          \begingroup \bm:counta=\bm:countd
                          \divide\bm:counta by "1000
                              \begingroup
                              \multiply\bm:counta by "1000
                              \global\advance\bm:countd by
                                                      -\the\bm:counta
                              \endgroup
                          \bold:mathrecode
                          \multiply\bm:counta by "1000
                              \begingroup
                              \bm:counta=\bm:countd
                              \bold:mathrecode
                              \global\bm:countd=\bm:counta
                              \endgroup
                          \global\advance\bm:countd by \the\bm:counta
                          \endgroup
                      \else\ifboldwarning\bm:message{\the\bm:tok\space
                        is not a \string#2.^^JDoing the obvious thing..}%
                           \fi
                      \bold:mathrecode
                      \bm:countd=\bm:counta
                      \multiply\bm:counta by "1000
                      \advance\bm:countd by \the\bm:counta
                      \fi
                      \advance\bm:countd by \expandafter"\thehex
                                                        \bm:countc000000
                                                %D HOLDS THE NEW DELCODE
                      \expandafter\xdef\csname bold\string#1\endcsname
                               {#2\the\bm:countd}%
          \else\bold:mathrecode
               \advance\bm:counta by \expandafter"\thehex\bm:countc 000
                                                %A HOLDS THE NEW
					        %MATHCODE
               \ifx#2\mathchar
                   \global\expandafter\mathchardef\csname
                                bold\string#1\endcsname=\the\bm:counta
               \else\expandafter\xdef\csname bold\string#1\endcsname
                               {#2\the\bm:counta}%
               \fi
          \fi
    \fi
}%
\def\bold:mathrecode{%ATTEMPTS TO COMPUTE A BOLD MATHCODE.  ACTS ON
		     %COUNTER A
      \bm:countb=\bm:counta
      \divide\bm:countb by "100%                B HOLDS THE FAMILY DIGIT
      \advance\bm:counta by -\expandafter"\thehex\bm:countb 00
%                                                 A HOLDS THE CHARCODE
%     ROUTINE FOR REASSIGNMENT OF FAMILIES
      \ifcase\the\bm:countb
              \bm:countb=\the\bffam%               FAM0 -> BFFAM
      \or     \bm:countb=\the\bm:bmitfam%          FAM1 -> BM:MITFAM
      \or     \bm:countb=\the\bm:bsyfam%           FAM2 -> BM:SYFAM
      \else   \ifboldwarning\ifbm:delim\bm:message{Lack of bold
                                        extension fonts means
                                        \string\bold\the\bm:tok\space may
                                        not be bold.}%             
                             \else\bm:message{Sorry, there just aren't the
                                   fonts for \string\bold\the\bm:tok.}%
                             \fi
              \fi 
      \fi
\advance\bm:counta by \expandafter"\thehex\bm:countb 00
                                             %A NOW HOLDS THE NEW MATHCODE
                    }%
\def\DeclareBoldMacro#1#2#3{%\DeclareBoldMacro{NAME}{TYPE}{CODE}
    \bm:counta=#3 \bm:countd=\bm:counta
    \ifnum\the\bm:counta>"7FFF 
          \divide\bm:counta by "1000
    \else
          \multiply\bm:countd by "1000
    \fi
    \bm:countc=\bm:counta
    \divide\bm:countc by "1000
    \advance\bm:countd by -"\thehex\bm:countc 000000
    \edef\bm:process{\noexpand\defboldsymbol{\noexpand#1}{\noexpand#2}%
                     {\the\bm:counta}{\the\bm:countd}}%
    \bm:process}%
%
%
%%%%%%%%%%%%%%%%%%%%%%%%%%%%%%%%%%%%%%%%%%%%%%%%%%%%%%%%%%%%%%%%%%%%%%%%%%
%MACRO FIXES
\DeclareBoldMacro{\{}{\delimiter}{"4266308}
\DeclareBoldMacro{\}}{\delimiter}{"5267309}
\DeclareBoldMacro{\langle}{\delimiter}{"426830A}
\DeclareBoldMacro{\rangle}{\delimiter}{"526930B}
\edef\FixLessThanGreaterThan{\noexpand\DeclareBoldMacro{<}%
{\noexpand\mathchar}{\the\mathcode`\<}%
\noexpand\DeclareBoldMacro{>}{\noexpand\mathchar}{\the\mathcode`\>}}%
\FixLessThanGreaterThan
\DeclareBoldMacro{\sqrt}{\radical}{"270370}
%%%%%%%%%%%%%%%%%%%%%%%%%%%%%%%%%%%%%%%%%%%%%%%%%%%%%%%%%%%%%%%%%%%%%%%%%%
\default
\def\boldspecials{\when\mathmit\use\mathbm:bmit;\when\mathcal\use\mathbm:bcal;\when\mathrm\use\mathbf;\when\mathsf\use\mathbm:bsf;\when\mathit\use\mathbm:bit;\when\mathtt\use\mathtt;\when\mathsl\use\mathbm:bsl;\when\default\use\default;}%
%%%%%%%%%%%%%%%%%%%%%%%%%%%%%%%%%%%%%%%%%%%%%%%%%%%%%%%%%%%%%%%%%%%%%%%%%%
\boldwarningtrue
\catcode`\:=12
%%%%%%%%%%%%%%%%%%%%%%%%%%%%%%%%%%%%%%%%%%%%%%%%%%%%%%%%%%%%%%%%%%%%%%%%

\magnification \magstep 1
\hsize 6 true in
\vsize 8.5 true in
\input amssym.def
\input amssym.tex
%\nopagenumbers
\openup 1 \jot
\centerline {\bf Collapsing Shells and the Isoperimetric Inequality for 
Black Holes}
\vskip 1.5 cm
\centerline {G.W. GIBBONS}
\centerline {D.A.M.T.P.}
\centerline {University of Cambridge}
\centerline {Silver Street}
\centerline {Cambridge CB3 9EW}
\centerline {U.K.}
\vskip 1.5cm
\centerline  {\bf ABSTRACT}
\tenrm
{\narrower \narrower \smallskip
Recent results of Trudinger on Isoperimetric
Inequalities for {\sl non-convex} bodies are applied to the 
gravitational collapse of a lightlike shell of matter
to form a black hole. Using some integral identities
for co-dimension two surfaces in Minkowski spacetime, the area $A$ of the apparent horizon is
shown to be bounded above in terms of the mass $M$ by the $16 \pi G^2 M^2$,
which is consistent with the Cosmic Censorship Hypothesis.
The results hold in four spacetime  dimensions and above.} 

\beginsection  0 Introduction

Some years ago Penrose suggested a 
means of producing a 
counter-example to the Cosmic Censorship Hypothesis by 
considering the collapse
at the speed of light of  
of a thin shell, $\cal N$, of matter with total mass $M$ [1]. During the collapse 
a closed marginally outer trapped outer 2-surface $T$ is formed
with area $A(T)$. Penrose pointed out that consistency with our
conventional ideas demands that 
$$
GM\ge \sqrt {A \over 16\pi},
\eqno(0.1)$$ 
with equality only in the spherically symmetric case.
This inequality, which is sometimes called the Penrose inequality,
is expected to hold for the ADM mass of a general asymptotically flat data 
set containing an apparent horizon.  
It has thus come to be of interest in its own right,
quite independently of its connection with Cosmic Censorship.
Its general validity
would constitute a significant and geometrically elegant
strengthening of the Positive Mass Theorem. It has
an appealing  interpretation in terms of 
the variational characterization of the static black hole equilibrium
states. Finally, and more speculatively, one may attempt to extend
the notion of the Bekenstein-Hawking entropy $S$
for stationary event horizons
to the time dependent case 
$$
S(t) = { 1\over 4G} A(t)
\eqno(0.2)
$$
where now  $A(t)$ is the area of a cross section of the event
horizon, i.e of a black hole $B(t)$, at time $t$. One might then
attempt to bound $A(t)$ below by the area $A(t)$ of an apparent horizon
lying inside the black hole $B(t)$.

For these reasons, and because it actually coincides
with it in some particular cases, I have suggested  [2]
calling
the inequality (1)
the 
\lq isoperimetric inequality for black holes\rq.
This would not only indicate its intrinsic importance but puts it
in the more general context of the basic geometric
inequalities, such as the eponymous example,  which play
such a significant role in geometry and physics. Moreover
emphasising this connection might lead to the importation
into black hole theory of further useful ideas 
and techniques from global analysis.

The initial evidence for the isoperimetric inequality was
limited to  instructive particular examples
some of them reported in [1] and [3,4 ]. Since that time
more evidence has accumulated but we still lack a general proof.
It has also become clear that the conjecture
need not be restricted to four space-time dimensions. With 
the obvious 
appropriate adjustment of the factor $16 \pi$ to take into account
conventions about
the definition of the ADM mass and the areas of unit spheres
in higher dimensions the inequality should continue to hold.
In five spacetime dimensions this would lead to some interesting
properties of the euclidean gravitational action.
It is also possible to generalize the inquality
to incorporate a cosmological term.

Penrose took the interior $\cal M^-$ of the shell to be isometric to
the interior of a null hypersurface $\cal N$ in
flat  Minkowski spacetime ${\Bbb E}^{3,1}$.
This allows one to calculate both the mass $M$ and the area $A$
using Minkowski geometry.  In [3] I pointed out that 
if the marginally outer-trapped surface $T$ lay in a flat
spacelike hyperplane, then the inequality is  equivalent
to a well known inequality of Minkowski [5] relating the mean curvature
integral of a convex body $T$ in euclidean 3-space ${\Bbb E}^3$
to its area. In fact Minkowski's inequality  is,
in the case of a convex body, 
intimately connected with the standard isoperimetric inequality
for the volume enclosed by a closed surface of area $A(T)$. 
I also showed, by means of some
calculations  that the more general case in which $T$ did not lie
in a flat
spacelike hyperplane would also hold if Minkowski's inequality continued
to hold for a body in euclidean 3-space ${\Bbb E}^3$
whose mean curvature
$$
J=  \Bigl ( { 1\over R_1} + { 1\over R_2} \Bigr )
\eqno(0.3)$$
is non-negative. Here $R_1$ and $R_2$ are the principal radii of curvature,
both of which must be positive if the body is to be (strictly) convex.
In fact if $T$ lies in a flat hyperplane the conditions
of the problem demand that $T$ be convex but if $T$ does not lie
in a flat hyperplane then my procedure was to project $T$ othogonally 
onto an
arbitrarily chosen spacelike hyperplane in its past to give a 
projected 2-surface $\hat T$ and then to work  with $\hat T$.

My calculations showed that $\hat T$ must have non-positive mean
curvature and a few numerical calculations, especially with a catenoid
capped with two flat discs, encouraged me to believe that
indeed Minkowksi's inequality continued to hold but  all the 
proofs in the literature that I could find
at that time [6] made essential use of convexity and did not seem to generalize.
For that reason I did not publish my calculations.  

Subseqently Tod [7,8,9,10,11] made further progress. Following work by Penrose
he considered  the special case when $\cal N$ is the past light cone
of a point in Minkowski spacetime ${\Bbb E}^{3,1}$.
Remarkably he was able to show the equivalence in that case with another
example of the iso-perimetric inequality. 

Recently Robert Bartnik has drawn my attention to 
some new work of Trudinger [12]
in which he establishes a large number of 
strenghthened iso-perimetric inequalities. Among them
is the case I needed. Thus we now have the general proof for the example
considered originally by Penrose. 
Moreover Trudinger's work
is valid in any dimension and so encourages  one to generalize the 
the original setup to any spacetime dimension and 
check the inequality in that case. That is the purpose
of this present paper which thus includes my original unpublished
calculations as a special case.

The paper begins in section (1) with a description of the basic set up
in $n+1$ spacetime dimensions
and then in section (2) develops a convenient formalism for describing the 
integral differential
geometry of a spacelike $n-1$ surface $T$ embedded in
 $n+1$ dimensional Minkowski-spacetime ${\Bbb E}^{n,1}$.
In the original case of $n=3$ this is a fairly straightforward
adaption of a discussion in [13] for 2-surfaces embedded in 3-dimensional 
euclidean space ${\Bbb E}^3$. It is 
hoped that this formalism may prove useful for other purposes.
This machinery is used in section (3) to establish some useful
integral identities. In sections (4) and (5) we establish our basic result
and section (6) is a conclusion.

\beginsection 1 Collapsing Shells.

The situation envisaged by Penrose [1] which has been elaborated upon
in detail by Barrab\'es and Israel [14] is that of a thin shell of matter 
which collapses at the speed of light in an asymptotically flat spacetime $\cal M$.
The shell is an ingoing null hypersurface $\cal N$ and
the spacetime $\cal M$  is the union of the interior ${\cal M}^- $
and the exterior ${\cal M}^+$ of $\cal N$. 
The interior ${\cal M}^-$ is taken to be isometric to
a null hypersurface, which we also call ${\cal N}$, in flat
Minkowski spacetime
${\Bbb E}^{n,1}$. The exterior region ${\cal M}^+$ is not flat and contains
(unless the collapse is spherically symmetric) gravitational radiation.   
The shell $\cal N$ carries energy and momentum of the form
$$
T_{\mu \nu}= \epsilon n_\nu n_\mu 
\eqno(1.1)$$
with $n_\mu n^\mu=0$. In the limit that the shell becomes
indefinitely thin we have distributional source and the spacetime 
has a discontinuity across $\cal N$.

The world lines of the null matter are then taken to coincide with
the null geodesic generators $n^\alpha$  of the null hypersurface
$\cal N$ and it is conveninent to choose them to be affinely parameterized
so that
$$
n^\alpha\, _{;\beta}n^\beta=0. 
\eqno(1.2)$$
Because $\cal N$ is ingoing  the expansion should be negative, i.e.
$$
n^\alpha\, _{;\alpha} <0. 
\eqno(1.3)$$
Note that the divergence coincides with the expansion because
we are using an affine parmeterization. 

One now considers the outgoing null hypersurfaces $\cal L$ starting out from
an $(n-1)$-dimensional surface $S $ lying just outside
the shell $\cal N$. Let $l^\alpha$
be the geodesic generators of $\cal L$ which we can
also take to be affinely parameterized. These outgoing null geodesics
may be expected to escape to infinity if $S$ is well to the past but
eventually one would expect, as one moves it to the future along $\cal N$,
 that $S$ would become outer trapped,  i.e everywhere on $T$ one has, because we are using an affine parmeterization,  
$$
l^\alpha\, _{;\alpha} <0.
\eqno(1.4)$$

Since the outgoing null geodesics started out inside $\cal N$
which is isometric to a portion of 
Minkowski spacetime they must have started 
out diverging i.e with
$$
l^\alpha \, _{;\alpha} >0.
\eqno(1.5)$$
By applying the Raychaudhuri equation to a bundle of null geodesic
rays parallel to $l^\alpha$ one deduces that the change in the
expansion is proportional to the energy carried by the shell.  
Now imagine that there is a marginally trapped surface $T$
lying on $\cal N$, i.e. one for which everywhere on it
$$
l^\alpha \,_{;\alpha}=0. 
\eqno(1.6)$$
One deduces that the the energy carried by the shell is
proportional to expansion of the outward null normals
of $T$, regarded as embedded in Minkowski spacetime.
Moreover because the surface is marginally
trapped it cannot lie outside the event horizon 
and thus its  area $A(T)$, again regarded
as embedded in Minkowski spacetime, should provide
a lower bound for the final area of the event horizon.  
Note that it is only consistent to assume that the metric inside the shell
is flat if no points in ${\cal M}^-$ 
are in the timelike future of the collapsing shell.
This means that the null geodesic generators
of $\cal N$ can have no caustics or focal points
to the past of the marginally outer trapped surface $T$.
If it happens that $T$ lies in a flat spacelike hyperplane
then $T$ must therefore be convex. 

In this way one reduces the general relativity problem
to one about the geometry of surfaces in Minkowski spacetime.
Specifically one needs
to know the ratio of the
$1 \over n-1$ root
of the area $A(T)$ to  $ 1 \over n-2$ root  of the
the integral
$$
\int _T l^\alpha\,  _{;\alpha}dA. 
\eqno(1.7)$$
One does not really need the factor of proportionality
since this is determined by the spherically symmetric,
i.e. $SO(n)$-invariant,  case.
Therfore we shall not work it out explicitly here.

\beginsection 2 Co-dimension two surfaces in Minkowski spacetime.

We are now going to discuss the differential geometry
of an $n-1$ dimensional submanifold $T$  of $n+1$ dimensional
Minkowski spacetime. We are of course interested 
in the case when the surface $T$
 is spacelike but most of the general formalism  
goes through if the surface is timelike and would thus be applicable to
the motion of an $n-2$-brane. In fact it would not be difficult to generalize much of the formalism to the case of arbitrary $p$-branes.
The general description of embeddings is of course not new
but what is important for us and perhaps not as widely known is 
the  discussion of integral formulae.

An immersion, or in our case an embedding, $X: T\rightarrow {\Bbb E}^{n,1}$
of a co-dimension surface $T$ into Minkowski spacetime ${\Bbb E}^{n,1} $ 
in local coordinates is specified by the embedding functions
$
X^\alpha (q^a) $ where $(q^a)$, $a=1,\dots, n-1$
 are coordinates on $T$ and $ X^\alpha$, $\alpha =0, 1,\dots ,n$ are 
inertial coordinates on ${\Bbb E}^{n,1}$. In what follows we shall make
repeated and unacknowledged use of the fact that vectors in Minkowski
spacetime may be parallelly propagated unambigously.
The induced metric $g_{ab}$ on $T$ is given by
$$
g_{ab}= { \partial X^\alpha \over \partial q^a} \eta _{\alpha \beta} 
{ \partial X^\beta \over \partial q^b}, 
\eqno(2.1)$$
where $\eta_{\alpha \beta} = {\rm diag }( -1,1,\dots ,1)$ 
is the flat metric, and is assumed to be spacelike.  Therefore $T$ has two
future directed lightlike normals $l^\alpha$ and  $n^\beta$ 
with $\eta _{\alpha \beta } l^\alpha l ^\beta =0= \eta _{\alpha \beta } n^\alpha n ^\beta$ which are 
assumed outward and inward respectively  For convenience
we have partiallly fixed the normalization
by the condition   
$$
 l^\alpha n _\beta =-1
\eqno(2.2)$$
where indices have been lowered using the metric $\eta _{\alpha \beta}$.
In what follows we shall 
not,
for brevity, always  distinguish verbally between covariant or contravariant Minkowski tensors, although
the correct placing of greek indices in formulae will be strictly adhered
to.

Later we will fix the remaining freedom in the scaling of the 
of the null normals by introducing  an arbitrary constant
future directed timelike vector  $ t^\alpha$ 
 such that
$t^\alpha t_\alpha=-1$ and then  fix the scale of  the 
inward null normal
by $n^\alpha t_\alpha =-1$. If we define  $\gamma =-t^\alpha l_\alpha$
one has:

$$
t^\alpha = s^\alpha +\gamma n^\alpha + l^\alpha
\eqno(2.3)$$
where $s^\alpha$ is a spacelike vector tangent to $T$ and hence
orthogonal to the null normals $l^\alpha$ and $n^\alpha$.
Since
$$\gamma = { 1\over 2} ( 1+ s^\alpha s_\alpha),
\eqno(2.4)$$
one has $\gamma \ge { 1\over 2}$.

Now, regardless of whether we introduce $t^\alpha$ or not,
we may invert the pulled back metric $g_{ab}$ to give $g^{ab}$ and then 
push it forward to give a rank $(n-1)$ projection operator
$$
g^{\alpha \beta}= g^{ab} {\partial X^\alpha \over \partial q^a}{\partial 
X^\beta \over \partial q^b}= \eta ^{\alpha \beta} + l^\alpha n^\beta + n^\alpha l ^\beta.
\eqno(2.5)$$
The projection operator is idem-potent in the sense that
$$
g^{\alpha \beta}\,\eta _{\beta \gamma}  \,g^{\gamma \delta}= g^{\alpha \delta},
\eqno(2.6)$$
and its kernel is spanned by the two null normals:
$$
g^{\alpha \beta} l_{\beta} =0=g^{\alpha \beta} n_{\beta},
\eqno(2.7)$$
and thus using it  any Minkowski vector $F^\alpha$ may be projected onto $T$
$$
F^\alpha = {\overline F}^\alpha + n^\alpha (F^\beta n_\beta) + l^\alpha (F^\beta l_\beta).
\eqno(2.8)$$ 

The essence of our formalism is to regard extrinsic quantities
such as  $l^\alpha$ as functions, or in this case $(n+1)$-tuples 
of functions, on $T$. Therefore we  introduce
a derivative operator ${\cal D} ^\alpha$ acting on scalars 
and differentiating them along $T$ by
$$
{\cal D} ^\alpha= {\partial X^\alpha  \over \partial q^a } g^{ab} {\partial \over \partial q^b}= g^{\alpha \beta} \partial_\beta.
\eqno(2.9)
$$
It is not difficult to give $\cal D$ an invariant characterization
in terms of the pull-back under the embedding $X$ but since this does
not greatly expedite the calculations I shall not do so. Since $l^\alpha$ and $n^\alpha$ are normal to $T$ one has
$$
l_\alpha {\cal D} ^\alpha=0= n_\alpha {\cal D} ^\alpha.
\eqno(2.10)$$ 
Since $T$ has co-dimension 2 it has two second fundamental forms.
Using the two lightlike normals we may define
$$
L_{ab}= l_\alpha { \partial ^2 X^\alpha \over \partial q^a \partial q^b}=-{\partial X^\alpha \over \partial q^a} {\partial l_\alpha \over \partial q^b}
= -{\partial X^\alpha \over \partial q^a} {\partial X^\beta \over \partial q^b} \partial _\alpha l_\beta 
\eqno(2.11)$$
and 
$$
N_{ab}= n_\alpha { \partial ^2 X^\alpha \over \partial q^a \partial q^b}=-{\partial X^\alpha \over \partial q^a} {\partial n_\alpha \over \partial q^b}
= -{\partial X^\alpha \over \partial q^a} {\partial X^\beta \over \partial q^b} \partial _\alpha n_\beta. 
\eqno(2.12)$$

The two second fundamental forms $L_{ab}$ and $N_{ab}$  
are symmetric by virtue of the fact that
$$
{\partial X^\alpha \over \partial q^a} l_\alpha=0= {\partial X^\alpha \over \partial q^a} n_\alpha.
\eqno(2.13)$$

We also define their traces 
$$
\mu= -{ 1 \over2 } {\cal D}^\alpha n_\alpha= -{ 1 \over2 }g^{ab} N_{ab},
\eqno(2.14)$$
and
$$\rho ={ 1\over 2} {\cal D}^\alpha l_ \alpha= -{ 1 \over2 } g^{ab} L_{ab}.
\eqno(2.5)$$
As we have defined them the second fundamental forms depend only
on the behaviour of the null normals on the surface $T$. However in the 
problem we are considering  $l_\alpha$ and $n_\alpha$ are defined off 
the surface $T$ as the tangents to the null geodesic generators of the  
ingoing $\cal N$ and outgoing $\cal L$ null hypersurfaces through
$T$. In particular if we chose them both to be affinely paramaterized
then
$$
l^\beta \partial _\beta l^ \alpha =0= n^\beta \partial _\beta n^ \alpha.
\eqno(2.16)$$
It then follows that
$$
2\rho=  \partial _\alpha l^\alpha,
\eqno(2.17)$$
and 
$$
-2\mu=  \partial _\alpha n^\alpha.
\eqno(2.18)$$

The quantities $\rho$ and $-\mu$ may thus be regarded as the 
expansions of the outward and inward normals $l^\alpha$ and $n^\alpha$
respectively. In the usual case both $\mu$ and $\rho$ will be positive.
If $T$ is marginally outer trapped then $\rho$ will vanish.
If both vanish then the spacelike surface $T$ will be extremal
with  respect to the area functional.

\beginsection 3  Integral Formulae

We  have the following identity for any Minkowski vector $F^\alpha$
$${\cal D}_\alpha F^\alpha = {\cal D}^\alpha {\overline F}_\alpha -2 \rho ( F_\alpha n^\alpha ) + 2 \mu ( F_\alpha l^\alpha)
\eqno(3.1)$$
Now the first term on the righthand side is just the covariant
divergence of the projected vector ${\overline F}^\alpha$.
To check this  explicitly one writes
$$
{\overline F}^\alpha = {\partial X^\alpha \over \partial q^a} {\overline F}^a
\eqno(3.2)$$
and recalls that the Christoffel symbols of the metric $g_{ab}$ are given
by
$$
[ab,c]=  {\partial X^\alpha \over \partial q^c} \eta _{\alpha\beta}
{\partial^2 X ^\beta \over \partial q^a \partial q^b}. 
\eqno(3.3)$$ 
Thus if
$T$ is closed ( i.e. compact without boundary) then Stokes's theorem
on $T$ yields:
$$
\int_T ({\cal D^\alpha  }F_\alpha) dA= -2\int _T \rho (F_\alpha n^\alpha)dA +2\int _T \mu (F_\alpha l^\alpha)dA,
\eqno(3.4)$$
where $dA$ is the area element on $T$. This formula leads to some interesting integral identities.

 If we take $F^\alpha=X^\alpha$ we obtain
$$
A= -2\int _T \rho (X_\alpha n^\alpha)dA +2\int _T \mu (X_\alpha l^\alpha)dA.
\eqno(3.5)$$ 
If we choose for $F^\alpha$ the arbitrary constant
future directed timelike vector  $ t^\alpha$ 
introduced above 
we obtain
$$
\int _T \rho dA= \int _T \mu \gamma dA.
\eqno(3.6)$$
As an illustration we note that these  formulae 
yield a simple proof that Minkowski 
spacetime admits no closed
trapped or marginally trapped spacelike surfaces. By definition this
requires $\mu$ positive and $\rho$ non-positive. This can only 
happen if both $\rho $ and $\mu$ vanish which can only occur if the 
area of the spacelike surface $T$ also vanishes which is a contradiction. 
If we choose for $F^\alpha$ an arbitrary constant Minkowski vector
we obtain:
$$
\int _T  \mu l^\alpha dA=\int _T  \rho n^\alpha dA.
\eqno(3.7)$$
By taking $F^\alpha= G^{{\mu_1}\dots  \mu_p} C^\alpha \, _{{\mu_1} \dots \mu_p}$
where $  C_{\alpha {\mu_1} \dots {\mu_p }}$ is an arbitrary constant Minkowski
tensor we obtain
$$
\int _T  {\cal D}^\alpha G^{\mu_1\dots  \mu_p}dA
= -2 \int_T  \rho G^{\mu_1\dots  \mu_p} n^\alpha dA +2 \int _T 
G^{\mu_1\dots ,\mu_p} l^\alpha dA.
\eqno(3.8)$$
Taking $G^\beta = X^\beta$ and recalling that  ${\cal D}^\alpha X ^\beta= 
{\partial X^\alpha \over \partial q^a} g^{ab} {\partial X^\beta \over \partial q^b}$ is symmetric in $\alpha$ and $\beta$ we obtain
$$
\int _T \mu X^{[\mu} l^ {\nu]} dA  = \int _T \rho X^{[\mu} n^ {\nu]}dA.
\eqno(3.9)$$
 A physical interpretation of these formulae will be given shortly.

\beginsection  4 The case when $T$ lies in a hyperplane.

Clearly in this case we are on the familiar ground
of a hypersurface in $n$-dimensional euclidean space ${\Bbb E}^n$.  
If $n=3$ we recover the formalism developed in chapter XII
of Weatherburn [13]. The following description will
use the notation of three spatial dimensions
but the discussion generalizes 
immediately to higher dimensions.Thus I have followed Weatherburn in 
using  the symbol $J$ for  the divergence of the normal.

If $t^\alpha$ is the timelike normal to the hyperplane
and  one has $s^\alpha=0$ whence $\gamma={ 1\over 2}$ and
$$
t^\alpha= { 1 \over 2} n^\alpha + l^\alpha.
\eqno(4.1)$$
The surface $T$ has spacelike unit normal $\nu ^\alpha$ which is orthogonal
 to $t^\alpha$
$$
\nu ^\alpha = l^\alpha - { 1\over 2} n^\alpha.
\eqno(4.2)$$
Since $t^\alpha$ is constant we thus have $2\rho=\mu$, moreover
the operator ${\cal D}^\alpha$ is orthogonal to $t^\alpha$, i.e. $t^\alpha {\cal D}_\alpha =0$ and thus
$$
\rho={ 1\over 4} {\cal D}^\alpha \nu _\alpha.
\eqno(4.3)$$
where the greek index may now be taken to range from $1$ to $n$.
The right hand side equals one quarter the trace of the second fundamental form of
$T$ regarded as embedded in ${\Bbb E}^n$. If $n=3$ we have
$$
\rho ={ 1\over 4} J = {1\over 4} \Bigl ( { 1\over R_1} + {1 \over R_2} \Bigr )
\eqno(4.4)$$
where $R_1$ and $R_2$ are the principal radii of curvature, i.e. the 
inverses of the eigen values of the second fundamental form.   

The integral identity:

$$
\int _T  ( \mu l^\alpha-\rho n^\alpha)dA =0
\eqno(4.5)$$
reduces to the known vector identity:
$$
\int _T  dA J {\bold \nu }=0
\eqno(4.6)$$
and the indentity:

$$
\int ^T X^{[\alpha }( \mu l^{\alpha ]}-\rho n^{\alpha ]})dA=0
\eqno(4.7)$$
to the known vector identity:
$$
\int _T dA J {\bold x} \times {\bold \nu}.
\eqno(4.8)$$
These have the following physical interpretation. If the surface $T$
were a perfectly elastic shell, such as a soap film, the 
net force per unit area it exerts is
$$
\Bigl ( {1 \over R_1} + {1 \over R_2 } \Bigr ) {\bold \nu}. 
\eqno(4.9)$$
 The identities (4.6) and (4.8) are an expression of Newton's third law: the total force and the total moment of this force must vanish. 
The full relativistic formulae 
have a similar interpretation. One may regard the quantity
$$
P^\alpha = { 1\over 4 \pi} \int _T \rho n^\alpha dA
$$
as the total ingoing four momentum associated with the surface $T$. Note that
it is independent of the scaling freedom of $l^\alpha$ and $n^\alpha$.
Moreover from (3.7)  $P^\alpha$ is the same as the  
outgoing normal momentum, i.e.
$$
P^\alpha= { 1\over 4\pi}  \int _T \mu l^\alpha dA.
$$
Similarly
$$
M^{\mu \nu}= { 1 \over 2\pi} \int _T \rho X^{[\mu} n^{\nu]} dA 
$$
may be interpreted as the 
total ingoing 
relativistic angular momentum associated to $T$. 
By (3.9) this equals the  total outgoing 
relativistic angular momentum.

As noted above,  by the conditions of the physical problem
the surface $T$ can have no focal points in its past and therefore
in the present case $T$ must be a convex surface. 
The isoperimetric inequality now reduces to the original 
Minkowski inequality [5] for convex bodies:
$$
{ 1\over 8\pi} \int_T J dA  \ge \sqrt{ A \over 4 \pi}.
\eqno(4.10)$$

\beginsection 5 The case when $T$ does not lie in a hyperplane.

In this case one may arbitarilly choose a spacelike hyperplane
with unit normal $t^\alpha$ and project $T$ orthogonally onto it
by means of the projection operator:
$$
h_\alpha ^\beta = \delta _\alpha ^\beta + t_\alpha t^\beta.
\eqno(5.1)$$
The projected surface we call $\hat T$. Thus if $T$ is given
by $q^a \rightarrow (t(q^a), {\bold X}(q^a))$,  then  $\hat T$ 
is given by $q^a \rightarrow (0, {\bold X}(q^a))$. The unit normal
to $\hat T$ orthogonal to $t^\alpha$ must satisfy $t^\alpha {\hat \nu} _\alpha =0$ and ${\hat \nu}_\alpha h^{\alpha \beta} a_\beta=0$ for all vectors $a_\beta$ tangent to $T$ and is therefore given by
$$
{\hat \nu}^\alpha = { 1 \over \sqrt{2 \gamma}} ( l ^\alpha- \gamma n^\alpha ).
\eqno(5.2)$$
 
The area element $d{\hat A}$ of the projected surface $\hat T$ is less
than the area element $dA$ of the surface $T$. The induced metric ${\hat g}_{ab}$ on $\hat T$
is related to that on $T$, $g_{ab}$, by
$$
 {\hat g}_{ab}= g_{ab} + {\partial t \over \partial q^a} {\partial t \over \partial q^b}.
\eqno(5.3)$$
Thus
$$
{\rm det} \,{\hat g} _{ab} ={\rm det} \,{\hat g} _{ab}( 1+ g^{ab} {\partial t \over \partial q^a} {\partial t \over \partial q^b} ). 
\eqno(5.4)$$
Now ${\partial t \over \partial q^b} $ is the spatial projection $s^\alpha$
of the unit normal $t^\alpha$ onto $T$. One has
$$
1+ s_\alpha  s^\alpha = 2 \gamma
\eqno(5.5)$$
and therefore
$$
dA= \sqrt {{ 1 \over 2\gamma}} d {\hat A}.
\eqno(5.6)$$

We now evaluate the mean curvature ${\hat J}$ of the projected surface $\hat T$.
If ${\hat \partial}_\alpha= h_\alpha ^\beta \partial _\beta$ is the
purely spatial derivative orthogonal to $t^\alpha$
one has the standard result that 
$$
{\hat J}={\hat \partial}_\alpha {\hat \nu}^\alpha.
\eqno(5.7)$$
But
$$
{\hat \partial}_\alpha= \partial _\alpha + t_\alpha t^\beta \partial _\beta.
\eqno(5.8)$$
Using the fact that ${\hat \nu} ^\alpha t_\alpha=0$,
$$
{ \partial}_\alpha {\hat \nu} ^\alpha  + t_\alpha (t^\beta \partial _\beta {\hat \nu} ^\alpha)
= \partial _\alpha {\hat \nu}  ^\alpha - \nu _\alpha (t^\beta \partial _\beta t^\alpha).
\eqno(5.9)$$
But $t^\alpha$ is a constant vector and therefore
$$
{\hat J}= \partial _\alpha {\hat \nu}  ^\alpha= \partial _\alpha ({ 1 \over \sqrt {2 \gamma}} ( l ^\alpha- \gamma n^\alpha )). 
\eqno(5.10)$$
Therefore finally
$$
{ 1\over 2}  {\hat J}={ 1 \over \sqrt {2 \gamma}} \rho+ {\sqrt {\gamma \over 2}} \mu.
\eqno(5.11)$$
Note that $\hat J$ is necessarilly non-negative.
We may now us our integral identities and the relation (5.6) between 
$dA$ and $d{\hat A}$ 
to show that
$$
\int _T \rho dA = \int _T \gamma \mu dA = {1 \over 4} \int _{\hat T} 
{\hat J} d {\hat A}. 
\eqno(5.12)$$

We are now done because the ratio
$$
( 4 \int_T \rho dA)^ { 1\over n-2}   / A ^{ 1\over n-1}     
\eqno(5.13)$$
is clearly never less than the ratio
$$
(   \int_{\hat T}  {\hat J}  d {\hat A} )^ { 1\over n-2}   / {\hat A} ^{ 1\over n-1},
\eqno(5.14)$$
and this latter ratio is, by Trudinger's [10] strengthened form of the 
Minkowksi
inequality, never less than the value it takes for the standard round
embedding of the $(n-1)$-sphere.

I am grateful to Paul Tod for suggesting the following
formulation of the result. The energy $P^0$ of the shell
in the frame determined by  $t^\alpha$ is given by
$$
P^0= { 1 \over 4 \pi} \int_T \rho dA.
\eqno(5.15)$$
We have shown that in all frames
$$
P^{0} \ge \sqrt { A \over 16},
\eqno(5.16)$$
 
Thus $P^\alpha$ is future directed timelike and
$$
-P^\alpha P_\alpha \ge { A \over 16 \pi G^2}.
\eqno(5.17)$$

\beginsection 6 Conclusion

After some delay it is now clear that the answer
to Penrose's original question: \lq can one set up
a contradiction to Cosmic Censorship using collapsing
shells?\rq ~ is  definitely no. The reason is the existence of an
inequality which, links the isoperimetric
properties of black hole with the second law of thermodymamics.

An intriguing question for further study is whether there exist
further geometric inequalities which constrain gravitational collapse.
For charged bodies there is of coures a Bogomol'nyi bound [2]. Perhaps 
this has a geometrical interpretation for collapsing shells.
We also know that in four spacetime dimensions
black holes cannot have greater  angular momentum
than     
$$
 G^2 M^2. 
\eqno(6.1)$$
However this cannot be a general upper bound for all possible data sets
because two particles scattering against one another with very little 
energy
may carry  a large amount of angular momentum if their impact parameter
is sufficiently great. Nevertheless (6.1) suggests that it might be 
worthwhile investigating the angular momentum using  the methods of 
this paper. In fact there are some similarities with
a classical Regge inequality for the angular momentum of a string [15]. 
However while the Regge inequality holds in all dimensions
there is no upper bound like (6.1) for black holes in higher dimensions [16].
Presumably a Regge inequality holds for $p$-branes as well. Finally,
in connection with higher dimensions it is worth pointing out that 
Trudinger establishes inequalities for  integrals
of other  elmentary symmetric functions of the principal radii
of curvature. It would be interesting to know what, if any,
is their
physical significance.

\beginsection Acknowledgements

It is my pleasant duty to thank Robert Bartnik for bringing 
Trudinger's work
to my attention, Professor Trudinger for confirming that his results
do indeed include the strengthened form of Minkowski's inequality,
and Paul Tod for helpful conversations.

\beginsection References

\medskip \item {[1]} R Penrose {\sl Ann New York Acad Sci} {\bf 224} (1973) 125

\medskip \item {[2]} G W Gibbons in {\sl Global Riemannian Geometry} ed T J Wilmore and N J Hitchin ( New York: Ellis Horwood) (1984) 

\medskip \item {[3]} G W Gibbons {\sl Ph D Thesis} Cambridge University (1973)

\medskip \item {[4]} R Penrose  {\sl Seminar on Differential Geometry} ed S T Yau ( Princeton: Princeton University Press) (1982)

\medskip \item {[5]} H Minkowski {\sl  Math Ann} {\bf 57} (1903) 447

\medskip \item {[6]} W Blaschke {\sl Kreis und K\"ugel}  Chelsea, New York (1949)

\medskip \item {[7]} K P Tod {\sl Class Quant Grav} {\bf 2} (1985) L65-L68 

\medskip \item {[8]} K P Tod {\sl Class Quant Grav} {\bf 3} (1986) 1169-1189

\medskip \item {[9]} K P Tod {\sl Class Quant Grav} {\bf 6} (1989) 1159-1163

\medskip \item {[10]} K P Tod {\sl Class Quant Grav} {\bf 9} (1992)  1581-1591

\medskip \item {[1]} K P Tod in {\sl Recent Advances in General Relativity} ed A I Janis
and J R Porter ( Boston: Birkhauser) (1992)

\medskip \item {[12]} N S Trudinger {\sl Ann Inst Henri Poincare} {\bf 11} (1994) 411-425

\medskip \item {[13]} C E  Weatherburn  {\sl Differential geometry of three dimensions} (vol 1)
 ( Cambridge: Cambridge University Press) (1927)

\medskip \item {[14]} C Barrab\`es and W Israel {\sl Phys Rev} {\bf D 43} (1990) 1129-1142

\medskip \item {[15]} J Scherk {\sl Rev Mod Phys} {\bf 47} (1975) 123

\medskip \item {[16]} R C Myers and M J Perry {\sl Ann Phys (N Y) {\bf 172} (1986) 304 
\bye